\documentclass[conference]{IEEEtran}
\IEEEoverridecommandlockouts
\usepackage{cite}
\usepackage{amsmath,amssymb,amsfonts}
\usepackage{algorithmic}
\usepackage{graphicx}
\usepackage{textcomp}

\usepackage{graphicx}
\usepackage{cite}
\usepackage{latexsym}
\usepackage{verbatim}
\usepackage{amsmath}
\usepackage{amssymb}
\usepackage{theorem}
\usepackage{url}
\usepackage{times}
\usepackage{multirow}
\usepackage{enumerate}
\usepackage{xspace}
\usepackage{longtable}
\usepackage{threeparttable}
\usepackage{xcolor,colortbl}
\usepackage{algorithm}
\usepackage[shortlabels]{enumitem}
\usepackage[normalem]{ulem} 
\usepackage{bm}
\usepackage{csquotes}
\usepackage[T1]{fontenc}
\usepackage[utf8]{inputenc}
\usepackage{babel}
\usepackage[font=small,labelfont=bf]{caption}
\usepackage{caption}
\usepackage{graphicx}
\usepackage{booktabs}
\usepackage{soul}

\def\BibTeX{{\rm B\kern-.05em{\sc i\kern-.025em b}\kern-.08em
    T\kern-.1667em\lower.7ex\hbox{E}\kern-.125emX}}

\makeatletter
\def\endthebibliography{%
	\def\@noitemerr{\@latex@warning{Empty `thebibliography' environment}}%
	\endlist
}
\makeatother

\date{}
\usepackage[
top    = 0.75in,
bottom = 1in,
left   = 0.625in,
right  = 0.625in]{geometry}

\makeatletter
\newcommand{\linebreakand}{%
\end{@IEEEauthorhalign}
\hfill\mbox{}\par
\mbox{}\hfill\begin{@IEEEauthorhalign}
}

\begin{document}

\title{Counter Denial of Service for Next-Generation Networks within the Artificial Intelligence and Post-Quantum Era\\}



\author{
\IEEEauthorblockN{Saleh Darzi}
\IEEEauthorblockA{\textit{\small{Computer Science \& Engineering}} \\
\textit{University of South Florida}\\
Tampa, FL, USA\\
salehdarzi@usf.edu}
\and
\IEEEauthorblockN{Attila A. Yavuz}
\IEEEauthorblockA{\textit{\small{Computer Science \& Engineering}} \\
\textit{University of South Florida}\\
Tampa, FL, USA\\
attilaayavuz@usf.edu}
}

\maketitle

\begin{abstract}
Given the rise in cyber threats to networked systems, coupled with the proliferation of AI techniques and enhanced processing capabilities, Denial of Service (DoS) attacks are becoming increasingly sophisticated and easily executable. They target system availability, compromising entire systems without breaking underlying security protocols. Consequently, numerous studies have focused on preventing, detecting, and mitigating DoS attacks. However, state-of-the-art systematization efforts have limitations such as isolated DoS countermeasures, shortcomings of AI-based studies, and a lack of DoS integration features like privacy, anonymity, authentication, and transparency. Additionally, the emergence of quantum computers is a game changer for DoS from attack and defense perspectives, yet it has remained largely unexplored. This study aims to address these gaps by examining (counter)-DoS in the AI era while also considering post-quantum (PQ) security when it applies. We highlight the deficiencies in the current literature and provide insights into synergistic techniques to bridge these gaps. We explore AI mechanisms for DoS intrusion detection, evaluate cybersecurity properties in cutting-edge machine learning models, and analyze weaponized AI in the context of DoS. We also investigate collaborative and distributed counter-DoS frameworks via federated learning and blockchains. Finally, we assess proactive approaches such as honeypots, puzzles, and authentication schemes that can be integrated into next-generation network systems for DoS prevention and mitigation. 

\end{abstract}

\begin{IEEEkeywords}
Counter-DoS; Artificial intelligence (AI); post-quantum security; next-generation networks; deep learning.
\end{IEEEkeywords}



\section{Introduction}
\label{sec:Introduction}

The increasing prevalence of cyber-attacks on advanced network systems (e.g., 5G/6G, SDNs, wireless communication networks \cite{glas2012signal}, etc.), along with the rise in botnet and IoT applications, underscores the heightened significance and frequency of these security threats. Denial-of-Service (DoS) attacks, the most widespread and alarming among them, pose a direct threat to network infrastructure, target service availability, causing interruptions in resource access, potentially crippling an entire network without compromising the underlying security framework and at minimal cost \cite{kadri2023survey}. 
The advent of open-source tools, enhanced processing power, and widespread inexpensive devices has facilitated the evolution of DoS attacks into diverse and increasingly sophisticated forms.  These include large-scale Distributed DoS (DDoS) attacks and stealthy Low-rate DoS (LDoS) attacks that can evade conventional countermeasures \cite{bala2024ai}. 
Growing concern over DoS cyber threats and their profound impact on real-world applications has led to the deployment of various technologies for DoS prevention, detection, and mitigation. Numerous studies, primarily based on AI methods, focus on developing effective counter-DoS mechanisms.


\subsection{Limitations of Existing Systematization Efforts on DoS} 
Despite extensive studies into DoS attacks and countermeasures over the years, notable gaps remain regarding systematization efforts, as some discussed below:

\textit{Isolation of DoS Countermeasure Silos:}  
Most studies concentrate on singular aspects of DoS such as detection, mitigation, or prevention mechanisms individually, while they mainly provide a taxonomy of DoS attacks and present potential solutions within specific applications (such as SDNs \cite{karnani2023mitigation}, Delay Tolerant Networks \cite{chaudhari2022systematic}, and Blockchains \cite{wani2021distributed}.). They mainly address a single architecture (centralized or decentralized \cite{agrawal2022federated}), or a specific type of DoS attacks and often omit other variants (e.g., DDoS, Low-rate DoS \cite{zhijun2020low}). This isolated treatment limits their potential impact, and while synthesizing counter-DoS is beneficial, no survey has analyzed in depth the synergy of counter-DoS techniques or the challenges of combining them.


\textit{Gap on Synergies for DoS Countermeasures and Various Cyber-security Services:}   Existing systematization efforts mainly focus on DoS, addressing the availability aspect but overlooking other critical properties such as privacy, authentication, anonymity, and accountability. For instance, from the user's perspective, their privacy is of primary importance, and it may outweigh the benefits of participating in counter-DoS mechanisms unless some assurance is given.  Therefore, it is crucial to analyze these seemingly contradictory properties via privacy-enhanced counter-DoS mechanisms. An authentication DoS-resistant scheme applies to many real-world scenarios but may undermine user privacy and anonymity concerns. There is a significant gap in the systematization literature for counter-DoS methods and their intersections with critical cybersecurity techniques.

\textit{Lack of Post-Quantum Threat Analysis:} The advent of quantum computers fundamentally alters the cybersecurity landscape including DoS attacks and countermeasures. It poses a substantial threat to the long-term security of next-generation networks by rendering conventional cryptographic techniques vulnerable \cite{darzi2023envisioning}. Quantum capabilities have demonstrated advantages in breaking Proof of Work (PoW) and proactive DoS solutions, and they might impact ML/AI approaches. Moreover, they could enhance the efficiency and scope of DoS attacks. Despite the profound implications of the PQ threat on DoS and its countermeasures, existing systematization efforts have not yet captured this vital aspect. 


\textit{Need for Comprehensive AI-based DoS Coverage:} The studies on AI algorithms for DoS detection and mitigation, primarily focus on algorithm performance evaluation metrics \cite{sowmya2023comprehensive}. However, a vast majority of these studies lack comprehensive analysis of cutting-edge AI techniques (e.g., Generative AI, Large Language Models, Incremental Learning) and do not cover proactive approaches (e.g., AI-based tokens, puzzles, and honeypots). Moreover, there is limited analysis on weaponized AI, Adversarial ML, and AI-powered DoS attacks.

\begin{figure*}[h]
	\centering
	\includegraphics[scale=0.45]{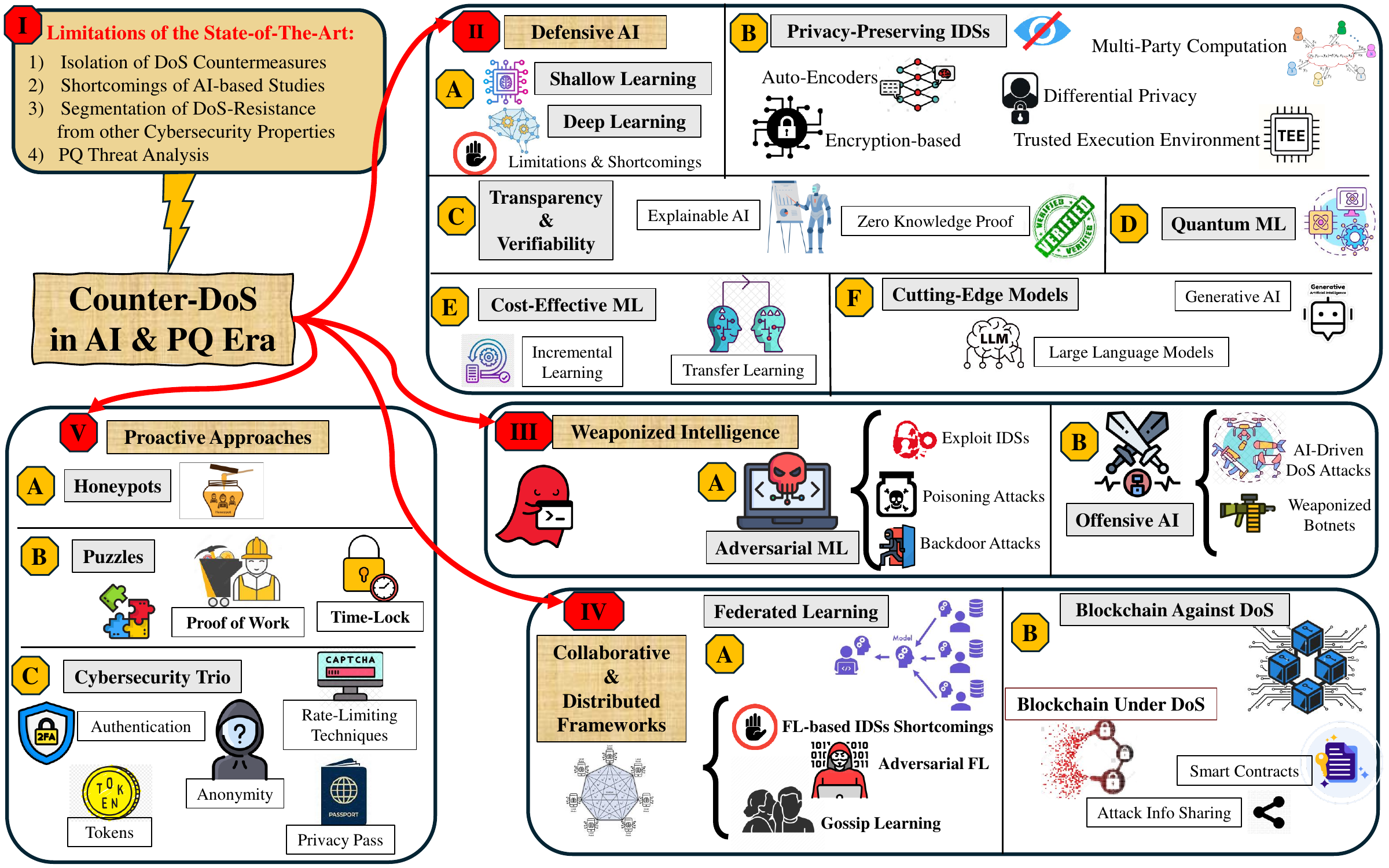}
	\caption{Taxonomy and Prospective on DoS and Counter-DoS Mechanisms in AI and PQ era}
	\label{fig:taxonomy}
	\vspace{-6mm}
\end{figure*}

\subsection{Our Contribution}
In this study,  we systematically investigate DoS and countermeasures in the context of AI and the PQ era for next-generation networks. Our analysis encompasses advanced AI models, collaborative and distributed frameworks, and proactive cryptographic approaches. We consider PQ security and the potential integration of DoS resistance with other cybersecurity services. We further outline our contributions below. 


$\bullet$ \ul{DoS and Countermeasures in the Post-Quantum Era:}  We consider direct quantum solutions such as Quantum Machine Learning (QML) and Post-Quantum Cryptography (PQC) to strengthen counter-DoS strategies and hybrid classical-PQ approaches. Additionally, we investigate the potential of PQC in tandem with privacy, authentication, and anonymity, when considered with DoS resistance.

$\bullet$ \ul{Synergy of Cyber-Security Properties with Counter-DoS:}  We juxtapose privacy with availability (equivalent to Counter-DoS), exploring anonymous solutions and authentication techniques while addressing transparency in decision-making processes, accountability of DoS adversaries, and the trustworthiness and robustness of counter-DoS systems. Navigating these conflicting needs, our objective is to prioritize cybersecurity properties alongside counter-DoS measures, exploring potential synergies among various techniques for a more effective and holistic solution.



$\bullet$ \ul{Examining Advanced/Weaponized-AI for DoS and Cryptographic Perspectives:} 
We explore the efficacy of AI-based solutions tailored for DoS in next-generation networks during the PQ era. We first delve into ML (shallow and deep learning) for Intrusion Detection Systems (IDSs), highlighting their limitations and potential advances. Next, we discuss AI-driven privacy-preserving countermeasures, transparency and verifiability in AI, unconditional security with QML, cost-effective ML optimizations, and the evaluation of cutting-edge AI models for countering DoS. Leveraging the cybersecurity features and PQ promises, we identify some shortcomings of these new AI techniques and provide potential insights for enhancements. Finally, we capture advanced weaponized AI for DoS, wherein the attack is enhanced via offensive AI.



$\bullet$ \ul{Examination of DoS in Collaborative and Distributed Frameworks:}  First, we explore federated learning (FL) as a scalable and privacy-preserving solution from a DoS perspective. Second, we examine blockchains as a potential architectural solution against DoS. Finally, we explore cybersecurity features and potential advancements for collaborative settings in the PQ era.


$\bullet$ \ul{Assessment of Proactive Approaches for DoS:}  We analyze DoS honeypots, puzzle-based approaches, and token-based cybersecurity trio (authentication, anonymization, and privacy). We analyze their advantages and limitations in both conventional and PQ secure contexts for integration into real-world applications.


{\em Taxonomy and Organization}: Figure \ref{fig:taxonomy} illustrates the taxonomy of our paper. Section II presents defensive AI-based mechanisms, while Section III elaborates on weaponized AI concerning DoS attacks. Section IV explores DoS strategies in collaborative and distributed settings. Section V navigates the proactive approaches for DoS prevention and mitigation. Section VI concludes the paper.

\section{Defensive AI: Fortifying Against DoS Attacks}
\label{sec:ids}

Recent advancements in AI technology have made it a focal point in cybersecurity, particularly for countering DoS attacks. However, AI is a double-edged sword. Here, we focus on its positive aspect as "Defensive AI". We evaluate AI's role in DoS detecting and mitigating, considering other crucial cybersecurity properties such as privacy, transparency, cost-effectiveness, and importantly, PQ security.

\subsection{State-of-the-Art ML Techniques for DoS-Specific IDSs}

In essence, ML-based DoS IDS functions as a sophisticated classification mechanism, utilizing statistical network transmission details to identify and block malicious entities. Their effectiveness depends on the types of DoS attacks, selected features, and network parameters. Despite extensive studies on ML-based IDSs for DoS attack detection, whether using shallow learning (SL) algorithms or deep learning (DL) models, there are still gaps in their utility and integration into real-world applications.   
The most common approach involves SL algorithms (e.g., SVM, Naive Bayes), which are typically supervised and more efficient than conventional counter-DoS measures (e.g., network-based filtering). However, they are not $100\%$ accurate and face issues such as misclassification, false positives, and the need for periodic re-evaluation. They require human intervention, are weak against zero-day (ZD) attacks, and typically need structured data, which is unrealistic and costly.   
Recent advancements are increasingly favoring DL models (e.g., CNN, RNN, GAN) to address these issues, leveraging their high accuracy, lower false alarm rates, and improved performance against ZD attacks \cite{habeeb2022network}. Despite the challenges posed by low-rate DoS attacks, which are hard to distinguish from normal network traffic patterns and require intricate network design for effective countermeasures, DL-based IDSs are capable of detecting these complexities. However, they are generally more complex, involve long training times, require high computational power, and are unsuitable for real-time networks. In summary, ML techniques (SL or DL) employed for DoS-specific IDSs have shortcomings that impede their full implementation in real-world scenarios. These issues and their potential solutions are discussed in the next subsections and summarized in TABLE I.

\begin{table}[tbh]
    \setlength{\tabcolsep}{2pt}
\begin{tabular}{|{c}|{l}|{l}|}\hline
\textbf{Issues} & \textbf{Limitations as Counter-DoS} & \textbf{Potential Solutions} \\\hline\hline
\multirow{3}{*}{Dataset} & Mostly Synthetic Datasets & Dataset Benchmarking\\
& Mostly Outdated Datasets  & Use Honeypots/Honeynets\\
& Lack of DoS Specification & Generative AI \\ \hline\hline
\multirow{3}{*}{Performance} & High Misclassification Rate & Train on Real-Time Network\\
& High False Alarms Rates & Use Advanced DL Models\\
& Inability to Detect ZD Attacks& Incremental Learning\\ \hline\hline
\multirow{3}{*}{Overhead} & Heavy Computational Costs& Online-Offline Structure\\
& Time-Consuming Labeling & Outsourcing Operations\\
& Long Training Time & Transfer Learning \\ \hline\hline
\multirow{3}{*}{Applicability} & Limited Scalability& Collaborative-IDS\\
& for Large-Scale Networks& Use Blockahins\\
& for IoT Environment & Federated Learning\\\hline\hline
\multirow{5}{*}{Missing}& Confidentiality \& Privacy & Privacy-Preserving-IDSs\\\cline{2-3}
& Transparency \& Trustworthiness & Explainable AI \\\cline{2-3}
& Verifiability \& Accountability & Zero-Knowledge Proofs\\\cline{2-3}
Cyber- & PQ Promises & Quantum-ML\\ \cline{2-3}
Security & Cost-effectiveness & Transfer Learning\\\cline{2-3}
Properties & Dynamicity & Incremental Learning \\\cline{2-3}
& Fully Autonomous & Large Language Models\\\cline{2-3}
& Fairness \& Robustness & Federated Learning \\\cline{2-3}
\hline
\end{tabular}
\caption{Limitations of ML Techniques for DoS-Specific IDSs}
\vspace{-6mm}
\label{tab:Limitaions}
\end{table}

%
%
\subsection{AI-Driven Privacy-Preserving DoS IDS Frameworks}
\label{subsec:PPIDS}

When addressing system cybersecurity, particularly regarding DoS attacks which often require extensive network data monitoring, prioritizing data confidentiality and user privacy is paramount. In practical terms, cybersecurity services are commonly outsourced to third parties for efficiency, cost-effectiveness, and seamless security, raising significant privacy concerns. This creates a clear conflict between the system's need for availability through counter-DoS mechanisms and the users' right to privacy and data protection. From users' perspectives, privacy considerations outweigh the need for service availability, making any defense mechanism lacking privacy prioritization impractical for real-world deployment \cite{grissa2016efficient}. As a result, the concept of Privacy-Preserving IDS (PPIDS) has emerged, leveraging various cryptographic approaches to address these concerns \cite{hesamifard2018privacy}.

%
%
\begin{table*}[tbh]
    \centering
    \setlength{\tabcolsep}{2.15pt}
    \begin{tabular}{||c||l||l||c||}\hline
        \textbf{PP-IDS} & \multicolumn{1}{c||}{\textbf{Advantages}} & \multicolumn{1}{c||}{\textbf{Limitations}} & \textbf{PQ Promises}\\\hline\hline
        \multirow{2}{*}{\textbf{AE}}& $\bullet$ Low Probability of Reverse-Engineering & $\bullet$ Introduce Delays in the IDS & \multirow{2}{*}{Unexplored}\\
        & $\bullet$ Suitable for Distributed Environments & $\bullet$ Unsuitable for Real-Time Networks & \\\hline
        \multirow{3}{*}{\textbf{HE/FE}} & $\bullet$ Full Confidentiality \& Privacy & $\bullet$ Substantial Computational Overhead & Lattice-based\\
        & $\bullet$ Require No Honest-But-Curious Entity& $\bullet$ Require Significant Computational Resources & \& Multivariate \\
        & $\bullet$ Applicable to Centralized \& Distributed Architectures& $\bullet$ Require Implementation \& Hardware Optimizations & Cryptography\\\hline
        \multirow{3}{*}{\textbf{MPC}}& $\bullet$ Model and Data Privacy & $\bullet$ Impose High Communication Burden & PQ-Secure Primitives,\\
        & $\bullet$ Low Computational Requirements & $\bullet$ Scalability Issues on Large-Scale Networks& Symmetric Cryptography,\\
        & $\bullet$ Applicable to Distributed and IoT Environments  & $\bullet$ High Bandwidth Requirements& \& Encryption Techniques\\\hline
        \multirow{2}{*}{\textbf{DP}}& $\bullet$ Preserve Individual Privacy & $\bullet$ Low Accuracy of The Detection Model & Quantum Noise\\
        &$\bullet$ Resistant to Inference \& Data Extraction Attacks & $\bullet$ Require an Honest-But-Curious Entity  & Quantum-DP \\ \hline
        \multirow{2}{*}{\textbf{TEE}}& $\bullet$ Preserve the Data and Code Privacy & $\bullet$ High Hardware Requirements \& Costs& Requires Employment of\\
        & $\bullet$ Hardware Security Guarantees & $\bullet$ Performance Issues and Delay & PQ-Secure Primitives in TEE\\\hline
    \end{tabular}
    \caption{DoS-Tailored Privacy-Preserving Intrusion Detection System Approaches}\vspace{-5.5mm}
    \label{tab:PPIDS}
\end{table*}

PPIDS can be implemented through several approaches, as demonstrated in TABLE \ref{tab:PPIDS}: 1) \textit{AutoEncoders (AE)} are unsupervised compression engines that preprocess raw data into an encoded form for ML classification, thereby preserving data privacy.
2) \textit{Encryption-based techniques}, such as Homomorphic Encryption (HE) and Functional Encryption (FE), enable training on encrypted data, while due to the substantial computational overhead, most approaches use pre-trained models for prediction or classification over encrypted inputs \cite{hesamifard2017cryptodl}.
3) \textit{Multi-party computation (MPC)} enables the secure distribution of a function (such as an ML model) on private inputs among parties without mutual trust \cite{sedghighadikolaei2024privacy, zhou2024secure}. 
4) \textit{Differential Privacy (DP)} involves adding controlled noise or making dataset-swapping modifications to preserve the statistical characteristics of the ML model while protecting individual data privacy \cite{mokry2021efficient}.
5) \textit{Trusted Execution Environment (TEE)} is a hardware-based approach that creates a secure and trusted environment for protecting ML data and code \cite{darzi2024pqc, hoang2020mose}.

\subsection{Transparency and Verifiability in DoS Defense}

The effects of selected DoS attack features on the accuracy of the IDS, along with the need for clear decision-making in AI, highlight the massive consequences of backdoors, which could lead to wrong predictions and undetected attacks. This undermines mitigation efforts and exposes systems to cyber threats. Thus, the transparency, interpretability, and trustworthiness of the employed AI-based IDSs are crucial for DoS detection-mitigation. Henceforth, there is a global effort to standardize trustworthy AI \cite{kaur2022trustworthy} where institutions like ISO and NIST, along with numerous countries are working to define trust in AI and address its associated issues.  
Additionally, in real-world scenarios where DoS defense is outsourced or pre-built tools are used, verifying the integrity of computations and AI models is also crucial. Zero-knowledge proofs (ZKP) serve as a complementary solution to privacy-enhancing frameworks, ensuring the correctness of model outputs, and detecting faulty behaviors, especially when the model owner might engage in malicious activities or use a poisoned model \cite{xing2023zero}. Despite the importance of verifiability in AI-based IDSs and the availability of various PQ ZKP schemes, few studies address this feature, and none focus on their application in DoS defense.

Furthermore, most AI-based IDSs are trained to detect anomalies mainly with binary classification but often lack justification, explanation, or confidence regarding the attack itself. Clear distinguishing factors are essential for integrating IDSs into real-world applications. Trust is built on understanding, necessitating knowledge of AI decision-making for full confidence. Explainable AI (XAI) addresses this by offering clarified decision-making and explanations of local and global model behaviors, contrasting with black-box AI systems \cite{arreche2024xai}. XAI, primarily uses game theory and XAI graphs to elucidate model operations, providing transparency, trustworthiness, and interpretability in next-gen networks.

However, despite XAI fostering trust in the utilized tools, it can suffer from lower accuracy and performance in DoS detection-mitigation. In this context, there are two potential future approaches for improving XAI: 1) Model-Specific XAI: Tailoring XAI specifically for DoS detection-mitigation to enhance accuracy. This approach requires training, implementing, and testing XAI on real-time networks rather than static datasets. 2) Model-Agnostic XAI (suitable for any ML model): Integration of model-agnostic XAI with conventional AI systems known for high DoS detection accuracy. This is a perfect fit for distributed architecture and can involve combining XAI at both local and global levels with distributed collaborative IDSs that develop a global model based on local training.  
Also, XAI for cyber-threat detection should be evaluated from a human perspective (e.g., security analysts) to enhance its utility and user-friendliness. 

%
%
\subsection{Next-Generation Defense: Quantum ML for DoS Attacks}
\label{subsec:quantumML}

From a PQ perspective, QML represents the optimal synergy of quantum computing and ML, offering quadratic to exponential speedup and addressing key performance-related issues in traditional ML\cite{biamonte2017quantum}. 
Concretely, given their parallelization and computational power, and considering the long training times and complex operations required on large datasets, QML is an ideal solution.  
In the context of DoS IDSs, proposed schemes either deploy Quantum-enhanced ML by executing classical data on quantum computers or use Quantum-inspired ML, where classical data and computations are optimized by quantum mechanics principles. These schemes are typically evaluated on Noisy Intermediate-Scale Quantum (NISQ) processors, which serve as an alternative to quantum computers \cite{payares2021quantum}. Additionally, various pure quantum DoS countermeasures, such as IDSs using quantum ML algorithms (e.g., Q-SVM, Q-CNN), are applicable to real-world (quantum) applications like Quantum Key Distribution (QKD) systems \cite{nicesio2023quantum}. However, these schemes are still in their early stages and require extensive study for practical implementation.


Despite the acceleration of processes, efficient handling of large datasets, and overall speedups in DoS detection via QML, its applicability, and practicality as an IDS are hindered by several issues: 
1) A shortage of effective encoding mechanisms for transferring network data into quantum states, impacts attack detection efficiency and model robustness. 
2) Lower accuracy levels compared to classical ML methods, high complexity in training computations, and substantial resource requirements (e.g., memory), making it not cost-effective.  
3) Performance weaknesses due to inherent noise and computational errors, leading to higher misclassification rates and weaker evaluation metrics. 
4) The primary role of QML in efficient DoS detection and mitigation is when using quantum data and datasets, which remain unexplored; without this, QML has no significant advantage over traditional ML.

All in all, QML as a DoS countermeasure is in its early stages, requiring substantial performance gaps to fill, and is hugely impacted by the improvements in the quantum computing domain to be applicable in real-world scenarios.


\subsection{Cost-effective ML Enhancements for DoS IDSs}
\subsubsection{\textbf{Transfer Learning (TL)}} Addressing the cost-effectiveness of deployed AI, TL is introduced to tackle challenges such as limited learning datasets, costly data sampling, expensive labeling operations, and the need for updating samples to cover new attack types \cite{rodriguez2022transfer}. TL is an advanced architectural solution that reuses a previously trained model, transferring the learned knowledge to new or updated tasks instead of starting from scratch. This approach involves a sequential or parallel use of multiple AI-based IDSs, offers performance advantages, and enhances the detection of new attack vectors by leveraging knowledge from different but related environments \cite{mahdavi2022itl}.   
Given the privacy sensitivity of some network data, privacy-preserving mechanisms such as encryption-based techniques (e.g., HE) or secure computation methods (e.g., MPC) should be incorporated at the design level of TL. However, despite these advantages, the application of TL in DoS-specific IDSs remains under-explored. 

\subsubsection{\textbf{Incremental Learning (IL)}} IL is another cost-effective solution that incrementally learns new features, enhancing the overall detection mechanism \cite{data2021t}. Given the rapid evolution of cyber threats and new variants of DoS in next-gen networks, continuous learning and IDS updates are essential. Updating already trained IDS on static data to capture new attack variants and features is challenging, making the dynamicity and adaptability crucial for real-world use. Sequentially training a model on multiple datasets can lead to "forgetting attacks," where the new model forgets previously learned knowledge, necessitating retraining from scratch, which is impractical and inefficient \cite{mahdavi2022itl, constantinides2019novel}. IL addresses these issues by using multiple AIs in a tree or hierarchical structure with pruning or retraining to cover new attack types and enhance accuracy for ZD attacks. However, IL requires more efficient design solutions combined with privacy-enhancing techniques, to realize its full potential in DoS IDSs.

\subsection{Cutting-Edge AI Models for DoS Detection-Mitigation}
\subsubsection{\textbf{Generative AI (GAI)}} Essentially, GAI models produce random instances of specific data classes. In cybersecurity, GAIs play a dual role: they can function as a stand-alone IDS, classifying benign, malicious, and fake behavior, or serve as an auxiliary tool to ML-based IDSs for dataset enhancement. Given that datasets are mostly outdated and of low quality, GAIs can generate labeled data, which is typically labor-intensive, error-prone, and raises privacy concerns when done by humans. Additionally, GAIs can create synthetic data to balance datasets, improving the efficiency of detection systems \cite{halvorsen2024applying}.  
For instance, producing synthetic data only tailored for various DoS attack variants, significantly improves DoS detection. Additionally, GAIs could be trained to serve as an alert-generating mechanism in IDSs, suitable for integration into real-life applications \cite{alwahedi2024machine}. However, the effectiveness of GAIs in DoS detection depends on the quality of synthetic data in identifying attacks, and in some cases, may not be cost-effective. Despite these limitations, GAIs show promise as auxiliary tools in counter-DoS mechanisms, particularly in addressing performance issues such as reducing false positives.

\subsubsection{\textbf{Large Language Models (LLMs)}} Recent advancements in LLMs, such as OpenAI's GPT-3.5 and GPT-4, have highlighted their efficiency in natural language processing (NLP) and their ability to understand non-natural languages (e.g., photos and videos) for various smart technologies \cite{li2024dollm}. Few studies have analyzed the use of LLMs with raw network flow data to identify DoS attacks. These studies have shown that LLMs can achieve high accuracy in DoS detection and provide clarity in their decision-making.  
For example, OpenAI's GPT models have demonstrated over $90\%$ accuracy in DDoS detection with training on only few examples, showcasing their superiority in certain scenarios \cite{li2024dollm}. Note that LLMs are as effective as other AI-based DoS-specific IDSs with balanced datasets and far superior in handling imbalanced classifications or ZD attacks. However, their significant computational overhead and inference latency limit their effectiveness in real-time detection. Further advancements in designing LLMs for delay-sensitive applications are necessary to overcome these limitations.

%
%

%
%
\section{Weaponized Intelligence: \\The New Age of DoS Cyber Threats}

On the offensive side, AI can be weaponized in two ways: "Adversarial AI," which involves exploiting AI mechanisms to maliciously infiltrate the DoS defense system, and "Offensive AI," which is employed to launch DoS attacks.

\vspace{-1mm}\subsection{Adversarial ML for DoS and Counter-DoS Measures}

Identifying vulnerabilities in IDS and exploiting loopholes in AI systems, enables attackers to evade detection and execute DoS attacks. Concretely, given the sensitivity of AI performance to inputs during training and prediction phases, Adversarial Machine Learning (AML) leverages small perturbations, known as Adversarial Examples (AEs), to cause misclassification, misprediction, or incorrect results \cite{zhang2022adversarial}. By utilizing algorithms such as Generative Adversarial Networks (GANs) \cite{piplai2020nattack}, Dos-tailored AML examines network traffic to identify potential features that can bypass detection. In essence, AML targets system robustness and the integrity of IDSs, aiming to reduce accuracy and increase false alarm rates through misclassifications. Consequently, common poisoning and backdoor attacks fall under AML, where adversaries contaminate training data and models to bypass certain DoS attacks \cite{gao2020backdoor}. Even DL-based IDSs, despite their complexity and accuracy, are vulnerable to backdoor attacks. 

Typically, AML can significantly reduce the true positive accuracy of DoS IDS, for example, from over 90\% to below 50\%, and in some cases, achieve a 100\% success rate for DoS misclassification with minimal perturbation or queries \cite{alotaibi2023adversarial}. Even when applied in a black-box manner (with no or limited knowledge of the underlying ML), AML can still maintain a high success rate for DoS attacks. However, most AML schemes are proposed in white-box scenarios (with full knowledge of the ML and its features), which is less realistic for IDSs where attackers typically have limited capabilities and knowledge. 
Note that AML creates AEs, addressing only predefined attacks in the datasets and not zero-day DoS attacks. 
However, most countermeasures against AML are limited by high computational demands, reducing defense accuracy and performance, and often only mitigating the severity of attacks rather than fully preventing them. These countermeasures are also not strictly applicable to DoS IDSs. Given the high success rate of AML attacks and the lack of comprehensive mitigation strategies, adversarial robustness and resiliency are crucial, especially as AI-based IDS are integrated into next-generation network systems.  
Even in the quantum era, the inherent noise in current NISQ or quantum computers, along with data transfer from classical to quantum states, makes QML highly susceptible to disruptions and adversarial techniques (Quantum-AML) \cite{edwards2020quantum}. This results in increased misclassifications and reduced robustness. Therefore, there is an urgent need for DoS-tailored IDS with resilience against AE and AML attacks.


\vspace{-1.5mm}\subsection{DoS Attacks Powered by Offensive AI}
Despite their significant threats, the scope of Offensive AI (OAI) DoS attacks and potential countermeasures remain largely unexplored. OAI has many applications, including military, autonomous weapon systems (e.g., drones, guns, robots), and cyber weapons \cite{aiyanyo2020systematic}. In the context of DoS, AI is utilized in AI-driven DoS/DDoS attacks, AI-powered malware, and weaponized botnets \cite{mirsky2023threat}. Essentially, OAI serves as a DoS cyber weapon in real-world applications with the following characteristics: 1) OAI enables DoS attacks on a larger scale and spreads the attack faster. 2) By monitoring network traffic, OAI facilitates the discovery of vulnerabilities for DoS exploitation. 3) OAI automates DoS attacks without human intervention and can identify new attack vectors. 4) OAI is adaptive, learning from adversaries' previous mistakes to evade detection and bypass defense mechanisms. 

OAI is typically employed to generate synthetic network data (adversarial traffic), simulating DoS attacks and increasing the fault rate and false alarms of IDS, thereby compromising system robustness. Ironically, while AI-based IDSs monitor network traffic to detect abnormal behaviors, the same algorithms can be repurposed to identify vulnerabilities and conduct DoS attacks. 
Moreover, OAI becomes particularly dangerous when combined with other tools. For instance, OAI can enhance botnets, making them more adaptive to their environments and resilient to countermeasures, thus enabling adaptive DoS attacks. The convergence of OAI, IoT, and botnets can lead to the "Botnet of Things," revolutionizing DDoS attacks with potentially devastating consequences \cite{malatji2023offensive}. Additionally, OAI's ability to mimic human behavior can undermine proactive security measures (e.g., CAPTCHA \cite{nakatsuka2021cacti}), rendering them ineffective.  
Given their characteristics and potential, DoS attacks with weaponized AI are significantly more sophisticated, efficient, and devastating compared to conventional attacks. Traditional security measures are insufficient to detect and defend against these attacks. Therefore, there is a growing need for comprehensive defense methods that incorporate AI along with advanced cryptographic techniques to provide robust protection.


\section{Enhancing DoS Defense Through Collaborative and Distributed Frameworks}
\label{sec:Distributed}
This section evaluates collaborative IDSs and distributed approaches including federated learning (FL) and blockchains. 


\subsection{DoS Detection-Mitigation with  Federated Learning}

In essence, FL is an architectural, collaborative, and scalable solution for ML that emphasizes security and data privacy. In FL, entities do not need to share their raw data to build a model, making it suitable for privacy-sensitive technologies (e.g., healthcare systems, finance), as well as applications with distributed networks (e.g., edge-computing, smart cities) \cite{gao2020end}. 
Recent studies have focused on FL-based DoS defense for its advanced security, privacy preservation, and architectural benefits. By incorporating high data heterogeneity with diverse samples, FL achieves higher model accuracy and improved DoS detection, outperforming traditional DL models \cite{qiao2024transitioning}.


Besides high privacy guarantees, this approach also reduces data transfer, minimizing security breaches during the global model-building process. It distributes the computational burden across numerous edge entities and alleviates the communication overhead of large-scale models to address the efficiency issues of AI-based IDSs and is suitable for scenarios with constraint bandwidth \cite{alazab2023enhancing}. 
FL-based IDSs are cost-effective, as they do not require specific infrastructure, hardware, or maintenance, unlike centralized MLs. They achieve fault tolerance and robustness since the framework continues to function even if a device fails or undergoes maintenance, which is particularly useful for DoS detection in large-scale networks. Moreover, they are trained in real-time networks and continuously learn from edge devices, enhancing dynamicity and adaptability. 
Also, different IDSs or companies, even from various countries, can participate as nodes in this framework to train a global DoS detection model without sharing sensitive data \cite{alazab2023enhancing}.

\subsubsection{\textbf{FL-based IDSs Shortcomings}}
Since FL is not fully distributed, its disadvantages stem from the centralized aggregation process, which is vulnerable to single points of failure, requires a trustworthy entity, and compromises full data confidentiality unless secure aggregation is implemented \cite{qiao2024transitioning}. This can be achieved through various methods: 1) \textit{Federated Averaging:} The most common method, though potentially biased towards entities with more updates and characterized by slow convergence. 2) \textit{Differential Privacy:} Noise can be added locally/globally during each round, but this method often involves a trade-off between DoS detection accuracy and data privacy. 3) \textit{Secure Aggregation Techniques:} Utilizing PPIDSs in distributed settings, typically based on PQ-secure MPC and encryption methods.   
Given that local training reveals sensitive information about individual clients, these techniques are highly recommended for privacy-sensitive DoS defense but are computationally intensive or add significant communication costs. 4) \textit{Federated Transfer Learning (FTL):} A new concept combining FL and TL, offering collaborative advantages across various IDSs. FTL provides more accurate DoS detection while preserving privacy but is limited to networks with similar data distributions, reducing its utility. 

Most FL-based IDSs rely on supervised algorithms needing labeled samples, but the scarcity of suitable datasets limits their effectiveness. Imbalanced data results in lower detection performance, reduced reliability, and inability to detect ZD attacks. In embedded environments, where device limitations are poorly represented and some users cannot contribute labeled samples, the applicability of FL for DoS defense is further restricted. Thus, developing practical FL-based IDSs for resource-constrained environments remains a challenge. 
Moreover, inefficient aggregation mechanisms contribute to delays in responding to DoS, thereby limiting their effectiveness in real-time response. As with centralized models, maintaining transparency and verifiability is crucial. Blockchain-based mechanisms can be utilized to incentivize client contributions and ensure operational transparency among nodes. 

\subsubsection{\textbf{Adversarial-FL}}
FL-based IDSs are not only vulnerable to AML attacks like central ML systems but are even more susceptible due to their distributed nature and the collaborative training process involving potentially malicious clients or aggregators. This vulnerability exposes them to attacks such as model inversion, backdoor, and poisoning attacks, where malicious entities can inject harmful data and tasks into the training set during each round \cite{zhang2023delving}. These actions can compromise the integrity of the FL model, leading to failure in detecting DoS attacks. Therefore, ensuring the integrity of FL models requires rigorous client selection and verification processes, necessitating further research and attention.

\subsubsection{\textbf{Gossip Learning (GL)}}
GL operates as a fully decentralized approach, facilitating local training and data storage with direct node-to-node model aggregation, thus eliminating the need for an honest-but-curious aggregator. It provides a cost-effective and robust scalability solution compared to FL, particularly suitable for scenarios where finding a trusted aggregator is challenging \cite{hegedHus2019gossip}.  
Despite its advantages, existing GL-based IDSs are generally not tailored specifically for DoS attacks and introduce delays in aggregation, which diminishes its effectiveness in real-time DoS detection compared to FL.  
Hence, FL continues to outperform GL in DoS detection across various evaluation metrics, including accuracy and false alarms. A promising future direction could involve synergizing FL and GL to create a holistic solution for DoS defense with real-time utilization.

\subsubsection{\textbf{Quantum-FL (QFL)}}
The concept of QFL aims to improve operational efficiency through quadratic to exponential enhancements and parallelization in decentralized structures. While QFL shows promise as a synergy between FL and quantum computing, there are currently no purely Quantum-empowered FL-based IDS systems specifically tailored for DoS attacks \cite{zhang2022federated}. The practical implementation of true QFL is hindered by the current state of quantum computers and their hardware, which are not yet widely available at edge nodes. However, hybrid quantum-classical FL approaches have been explored for DoS detection, leveraging the strengths of both worlds that are feasible with current equipment. It is noteworthy that existing QFL-based IDS systems are primarily supervised, which may not align well with real-world datasets lacking proper labeling. Thus, exploring QFL with unsupervised algorithms holds potential for further development. 

Another approach, contrasting with QFL, is blind quantum computation. This method allows computations to be delegated to an untrusted quantum server without revealing sensitive information such as input, output, or computation details 
\cite{qiao2024transitioning}. Looking ahead to the PQ threat, an optimal strategy involves leveraging low-scale quantum computers and NISQ systems alongside conventional FL techniques. Specifically, by employing classical algorithms at the network edge and utilizing a quantum-capable aggregator, we can enhance real-time DoS mitigation, facilitating a seamless transition to the PQ era with broad applicability across various real-world environments, especially those with limited constraints. 

%
%
\subsection{Blockchain-enabled Strategies against DoS Threat}
\label{subsec:blockchain}
In the context of the DoS threat, blockchain is susceptible to blockchain-based DoS (BDoS) attacks, yet it also serves as an architectural, collaborative, and scalable solution for DoS mitigation \cite{wani2021distributed}. Specifically, despite the political and economic incentives behind attacking cryptocurrencies and targeting their PoW constructs, blockchains can still be employed in DoS countermeasures for large-scale applications while addressing centralization issues.   
They are particularly valuable in scenarios where multiple IDSs need to communicate without mutual trust and are reluctant to disclose sensitive data yet seek to strengthen their security measures. Blockchain represents one of the few counter-DoS approaches suitable for applications with distributed networks where conventional centralized DoS defense mechanisms are not viable \cite{wani2021distributed}. Moreover, in some centralized DoS countermeasures, the defense mechanism itself becomes a new target for DoS attacks, highlighting blockchain's effectiveness as a solution.

The core novelty in blockchain-based DoS mitigation lies in its collaborative approach to attack-information sharing, trustworthy record-keeping for DoS blacklist, and increasing the cost of DoS attacks through consensus mechanisms (e.g., PoW, Proof of Stake). Blockchains not only provide accountability by penalizing malicious nodes but also serve as a verification mechanism for post-DoS analysis due to their immutable public ledger validated by a network of nodes. This is primarily achieved via smart contracts, which are agreed-upon codes executed by blockchain nodes to signal DoS detection and mitigation across the network \cite{truong2023metacids}.  However, blockchain-based methods are typically not standalone DoS defenses. They are usually integrated with AI-based DoS detection techniques, providing a transparent attack-sharing mechanism and keeping model updates in the training phase.

However, blockchain-based solutions have limitations in real-time DoS protection due to extended transmission and confirmation periods. Their interoperability is also lacking, as they are often bound to specific applications (e.g., IoT, edge computing, SDNs), requiring modifications and architectural support. This limits their practicality and flexibility as comprehensive solutions \cite{chaganti2023survey}. 
Most blockchain countermeasures still require human intervention, while the vulnerability of the underlying smart contracts to DoS attacks and the employment of PQ secure blockchains in DoS mitigation systems, although promising, remain unexplored. 
The ideal synergy in distributed settings combines FL with blockchains, leveraging the strengths of both to enhance detection accuracy, offer advanced privacy, and provide identity management through anonymization or pseudonymity \cite{truong2023metacids}. 
\section{Proactive Defense Against DoS: Leveraging Puzzles, Tokens, and Credentials}
\label{sec:puzzle}\vspace{-1mm}


%
%
\subsection{Honeypots for DoS Defense}
A honeypot is a proactive DoS defense mechanism, functioning as a software-based vulnerable machine designed to attract attacks (e.g., malware, malicious accesses) in computing systems \cite{vishwakarma2019honeypot}. It creates a fake environment to lure DoS attackers, capturing and reporting them to the network administrator. Integrated into the service machine, the honeypot is an independent component that makes it difficult for even advanced attackers to avoid the trap.  
They can be deployed in both centralized and distributed settings, providing an advantage in new ZD exploits and DoS attack detection. 
Honeypots are among the few defense mechanisms that provide accountability in DoS defense.  
Recent advances have introduced AI-powered honeypots \cite{memos2020ai}, combining honeypots with AI-based IDSs to detect botnets and enhance ZD DoS attack detection. However, these schemes still face high false positive rates, bypassed attacks, high overhead, and vulnerability to backdoor attacks, necessitating thorough evaluation before full deployment.  
Additionally, quantum computation techniques can be used to design quantum honeypots to detect intrusions during unauthorized operations \cite{nagy2023quantum}. However, this is a nascent field, and commercial implementation is still far from practical.

%
%
\subsection{Puzzles as a Foundational Defense Mechanism}
Client Puzzle Protocols (CPPs) are mathematical prevention techniques that aim to involve users' resources, thereby increasing the cost of conducting DoS attacks. Depending on the puzzle type (e.g., computational, delaying, AI), the client must allocate resources (e.g., computational power, memory, bandwidth) to solve them and authorize the legitimacy of their request, making DoS attacks infeasible. These counter-DoS puzzles must provide certain properties, such as cost asymmetry (difficult for the client but efficient for the server), unforgeability, non-parallelization, and fairness \cite{ali2020foundations}.  

PoW-based techniques are a common and promising DoS prevention measure, but they are typically interactive and depend on the client's computational power (e.g., CPU). This approach lacks fairness for resource-constrained devices and is vulnerable to parallelization techniques and powerful entities like quantum computers. 
Another puzzle is the verifiable delay function (VDF) \cite{raikwar2021non}, a pseudorandom function (PRF) that produces a unique output and a verifiable proof after a specified number of steps.  
Given the difficulty of finding a trusted third party in real-world applications, and considering the parallelization weaknesses of other methods, time-lock puzzles (TLPs) \cite{afshar2023possibility}, also known as "encryption into the future," offer a solution. TLPs are based on encryption schemes where decryption is only possible after a certain amount of time with inherent sequential steps. They do not require trust in any third party, are resistant to parallelization, and have public verifiability. However, despite their high computational depth, better processors or quantum computers can still affect the validation process, impacting DoS prevention effectiveness. 
Also, they have several issues and can be bypassed or weakened, while imposing high costs on the servicing server. Expensive tasks (e.g., puzzle generation and distribution) can be outsourced to a trusted entity, though this is often impractical. Also, attackers can precompute puzzles to launch DoS attacks later or replay the same valid solution for access \cite{lai2023lattice}. 

The computational power and parallelization capabilities of quantum computers pose a significant threat to CPPs \cite{bard2022quantum}. Although most schemes are based on symmetric cryptography primitives, particularly hash functions, Grover's algorithm can reduce their security by half. This issue can be addressed with difficulty adjustments, but this leads to unfairness and impracticality in real-world networks. Consequently, several PQ-secure PoW mechanisms have been proposed, based on the hardness of solving lattice-based problems, isogenies, and multivariate quadratic equations. However, these PoWs are significantly more computationally expensive \cite{behnia2021lattice}.  
PQ-secure PoW and timed cryptography are in their early stages, and many proposed solutions are either broken or too computationally intensive for integration into next-gen networks. 

\subsection{The Cybersecurity Trio: Contradictory Needs of Privacy, Anonymity, and Authentication}

\subsubsection{\textbf{Rate-Limiting Tokens and Authentication Systems}}
Authentication systems are commonly used in challenge-response protocols with rate-limiting structures at their core to resist DoS attacks via access control.  
However, these schemes are often application-specific and provide DoS resistance only at a single network-layer (e.g., AI-based CAPTCHA for application layer) \cite{akama2023scrappy}.  
Authentication systems are typically employed alongside other counter-DoS mechanisms because the authentication mechanism itself can become a target for DoS attacks. These systems are either based on hardware tokens (e.g., client-side TEEs), or computational rate-assuring proofs \cite{nakatsuka2021cacti}. Thus, they either require specific hardware or lack PQ promises due to their underlying cryptographic problems, such as integer factorization and discrete logarithm problems. However, there is little to no research on PQ-secure authentication protocols for DoS resistance. 


\subsubsection{\textbf{Anonymization Techniques}}

Anonymity, which ensures identity protection via unlinkability and untraceability, often conflicts with the required authentication in counter-DoS techniques. 
Anonymization techniques address these needs in three ways: 1) \textit{Anonymous Tokens:} These limit the number of tokens, tags, or cryptographic keys a user can acquire. If a user exceeds the predefined threshold, they are deanonymized, deterring DoS attacks while ensuring accountability \cite{davidson2018privacy}. 2) \textit{ZKPs:} Users can assert their limited request rate without sharing sensitive information or attributes. 3) \textit{Group Signatures:} A group member can sign on behalf of the group, with the group admin able to trace the signer’s identity, thus providing accountability and anonymous DoS resistance \cite{nakatsuka2021cacti, yavuz2023beyond}. However, this approach lacks scalability for large-scale and distributed networks. Nevertheless, PQ-secure group signatures and ZKPs have not yet been utilized for DoS mitigation.

\subsubsection{\textbf{Privacy Pass for DoS-Resistance}}
Due to the issues with temporary puzzles, the computational burden of PoW, and poor user experience, especially for anonymous access,  
the concept of Privacy Pass (PP) has been introduced \cite{davidson2018privacy}. PP is a cryptographic password that serves as a privacy-preserving access control mechanism with a rate-limiting structure. It aims to synergize authentication, anonymity, and privacy. Besides ensuring the unforgeability and unlinkability of tokens, PP can provide access control without repeated identification processes \cite{kakvi2023sok}.  
PP constructions typically use variants of digital signatures, such as group, blind, attribute-based, and threshold signatures \cite{yavuz2023beyond, sedghighadikolaei2023comprehensive}, to prove the legitimacy of user interactions without human intervention. Specifically, clients send a commitment of their attributes (e.g., nonce, timestamp, number of access times) to the issuer. The issuer generates a signature on the commitment and sends it back. To redeem the token, the user employs a ZKP on the signature for the given attributes \cite{kakvi2023sok}. 
In the context of DoS attacks, if an attacker is blocked or unable to obtain tokens, it will realize it has been detected and can change its identification or pseudo-ID. PP can enhance DoS mitigation by encoding a validity bit into its construction, issuing invalid credentials to clients with a bad reputation, rendering them ineffective \cite{kreuter2020anonymous}.  
However, PP constructions are limited in scalability, allowing only one credential to be obtained at a time, which is insufficient for large-scale networks. Post-quantum PP schemes based on standardized lattice-based signatures, isogenies, code-based, and multivariate cryptographic problems \cite{hauri2024post} are functional but incur high computational costs and require significant bandwidth. 
Thus, PQ-secure PPs are still impractical for DoS defense in next-gen networks and require further development.

\section{Conclusion}
With the rise in cyber threats to real-world applications and network services, DoS attacks have become more sophisticated and easier to execute, making comprehensive countermeasures increasingly challenging. This study addresses these gaps by examining counter-DoS mechanisms in the AI era, focusing on PQ security while integrating essential cybersecurity properties such as privacy, anonymity, authentication, and transparency. We highlighted deficiencies in the current literature, evaluated advanced AI models, analyzed weaponized AI, and explored collaborative distributed architectures. Additionally, we assessed proactive approaches such as honeypots and puzzles that could be integrated into next-gen network systems for DoS prevention and mitigation. Our vision aims to bridge the literature gap in the AI and PQ era, offering valuable insights for future research in DoS countermeasures.
\vspace{-1mm}

\bibliographystyle{IEEEtran}

\bibliography{Chapters/SalehRef}

\end{document}